\documentclass[%
 reprint,
superscriptaddress,
 amsmath,amssymb,
 aps,
]{revtex4-2}

\usepackage{graphicx}
\usepackage{dcolumn}
\usepackage{bm}

\usepackage{comment}
\usepackage{mathtools}
\usepackage{amsthm}
\usepackage{xcolor}
\usepackage{bm}

\newcommand{\note}[1]{\textcolor{black}{#1}}

\newtheorem{theorem}{Theorem}

\newtheorem{corollary}{Corollary}[theorem]

\theoremstyle{definition}
\newtheorem*{example}{Example}

\begin{document}

\preprint{APS/123-QED}

\title{Theoretical constraints on \note{reciprocal and non-reciprocal}\\ many-body radiative heat transfer}

\author{Cheng Guo}
 \affiliation{Department of Applied Physics, Stanford University, Stanford, California 94305, USA}
 
\author{Shanhui Fan}
\email{shanhui@stanford.edu}
\affiliation{%
 Ginzton Laboratory and Department of Electrical Engineering, Stanford University, Stanford, California 94305, USA
}

\date{\today}

\begin{abstract}
We study the constraints on reciprocal and non-reciprocal  many-body radiative heat transfer imposed by symmetry  and the second law of thermodynamics. We show that the symmetry of such a many-body system in general forms a magnetic group, and the constraints of the magnetic group on the heat transfer can be derived using a generalized reciprocity theorem. We also  show that  
the second law of thermodynamics provides additional  constraints in the form of a nodal  conservation law of heat flow at equilibrium. As an  application, we provide a systematic approach to determine the existence of persistent heat current in arbitrary many-body systems. 
\end{abstract}

\maketitle

\section{Introduction}
\label{sec:intro}

Thermal radiation is important for both fundamental science and engineering applications~\cite{planck1991,rytov1989, chen2005, zhang2007, howell2016, fan2017}. The vast majority of the literature on radiative heat transfer assumes Lorentz reciprocity, which imposes strong constraint on radiative heat transfer~\cite{onsager1931a,onsager1931,casimir1945}. On the other hand, recently there have been significant progress in studying radiative heat transfer using non-reciprocal materials such as magnetooptical materials \cite{moncada-villa2015, zhu2014,silveirinha2017,abrahamekeroth2018,ott2018, ott2019a, zhao2019a,fan2020a} and magnetic Weyl semimetals \cite{zhao2020e,tsurimaki2020,ott2020a}. These studies have led to the discoveries of novel phenomena in nonreciprocal many-body radiative heat transfer such as persistent heat current \cite{zhu2016} and photon thermal Hall effect \cite{ben-abdallah2016, guo2019a,ott2020a}, which exemplifies the opportunities of exploring novel aspects of radiative heat transfer that can arise in complex reciprocal and nonreciprocal many-body systems \cite{ben-abdallah2011,khandekar2019b}.

In this paper, in order to provide theoretical guidance on the explorations of many-body radiative heat transfer, we consider the general theoretical constraints on such process. Certainly the heat transfer is constrained by the second law of thermodynamics. Moreover, for a non-reciprocal many-body system, where reciprocity is broken with either internal or external bias magnetic fields on at least some of the bodies, its symmetry consists of two classes of operations. The first class is the usual spatial operations, such as rotation and mirror operations, which transforms the  magnetic field bias on each of the body in the usual way of  pseudovectors. The second class consists of operations that  flip all the magnetic field bias in addition to the usual spatial operations. These two classes of operations together form the magnetic group of the many-body system. We show that the properties of many-body radiative heat transfer are strongly constrained by the structure of the magnetic group. The derivation in particular relies upon a generalized reciprocity theorem that relates the properties of two complementary systems. As an illustration of these theoretical constraints, we show these constraints can be used to identify   many-body non-reciprocal systems that do not  exhibit persistent heat current.

The rest of this paper is organized as follows. In Sec.~\ref{sec:theory} we provide the theory. In Sec.~\ref{sec:application} we apply our theory to determine the existence of persistent heat current in arbitrary many-body systems. We conclude in Sec.\ref{sec:conclusion}.

\section{Theory}
\label{sec:theory}

We consider a system consisting of $N$ bodies that exchange heat via radiation with each other and an environment. We label the environment (env) and the bodies as $\{ 0\equiv \mathrm{env}, 1, 2, ..., N\}$. In general, the system is an inhomogeneous  dispersive bianisotropic medium, which can be described by a $6\times 6$ constitutive matrix ${{C}}(\omega, \bm{r})$  \cite{tretyakov2002}:
\begin{equation}
    \label{eq:constitutive}
    \begin{pmatrix}
    \bm{D}\\
    \bm{B} 
    \end{pmatrix}
    = {{C}}
    \begin{pmatrix}
    \bm{E}\\
    \bm{H} 
    \end{pmatrix}
    =     \begin{pmatrix}
    {{\varepsilon}} & {{\zeta}} \\
    {{\eta}} & {{\mu}}
    \end{pmatrix}
    \begin{pmatrix}
    \bm{E}\\
    \bm{H} 
    \end{pmatrix},
\end{equation}
where ${{\varepsilon}}, {{\mu}}, {{\zeta}}, {{\eta}}$ are  $3\times3$ matrices of electric permittivity, magnetic permeability, electric-magnetic coupling strength, and magneto-electric coupling strength, respectively.   $\omega$ and $r$ denote the frequencies and the spatial coordinates, respectively. 
\begin{figure}[htbp]
    \centering
    \includegraphics[width = 0.6\columnwidth]{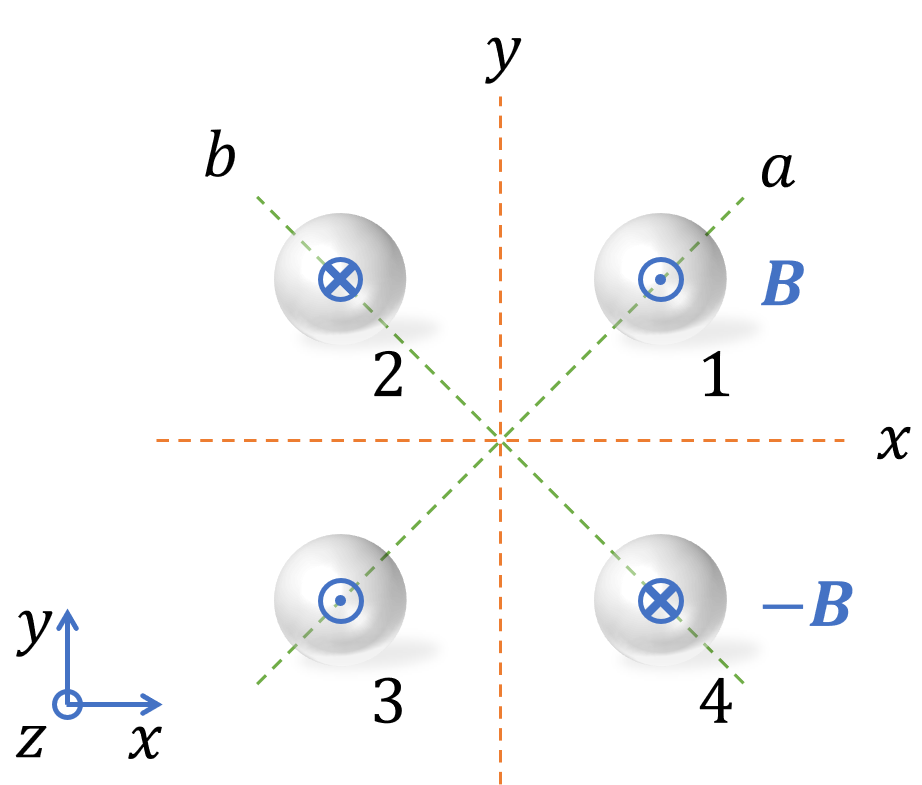}
    \caption{Schematic of a system consisting of four gyrotropic spheres that exchange heat via radiation with each other and an environment. The centers of the spheres form a square on the $x$-$y$ plane. There is a  magnetic field  along the $z$-direction with alternating signs in its distribution in the $x$-$y$ plane.}
    \label{fig:scheme}
\end{figure}

For radiative heat transfer, one considers the spectral heat flux to body $j$ due to thermal noise sources in body $i$ of temperature $T_i$:
\begin{equation}
    \label{eq:spectral_heat_flux}
    S_{i\rightarrow j}(\omega) = \frac{\Theta(\omega, T_i)}{2\pi} F_{i\rightarrow j}(\omega),
\end{equation}
where $\Theta(\omega, T_i) = \hbar \omega /[\exp{(\hbar \omega/k_BT)}-1]$, and $F_{i\rightarrow j}(\omega)$ denotes the temperature independent transmission coefficient from body $i$ to $j$. A general theory of many-body radiative heat transfer has been developed in Ref.~\cite{zhu2018} that allows to calculate $F_{i\rightarrow j}(\omega)$  given ${{C}}(\omega,\bm{r})$. 
All the directional transmission coefficients $F_{i\rightarrow j}(\omega)$ can be arranged into an \textit{exchange matrix}  $\mathcal{F}$ of dimension $(N+1)\times (N+1)$.
\begin{equation}
    \label{eq:adjacency_matrix}
    \mathcal{F} = 
    \begin{pmatrix}
        0 & F_{0\rightarrow1} & \ldots & F_{0\rightarrow N} \\
        F_{1\rightarrow 0} & 0 & \ldots & F_{1\rightarrow N} \\
        \vdots & \vdots & \ddots & \vdots \\
        F_{N\rightarrow 0} & F_{N\rightarrow 1} & \ldots & 0
    \end{pmatrix},
    \end{equation}
where we have suppressed the parameter $\omega$ for clarity. Our objective is to find the constraints imposed on $\mathcal{F}$ by the symmetry of the system as well as the second law of thermodynamics.

To illustrate the possible symmetry of these systems, we first consider a concrete example as shown in Fig.~\ref{fig:scheme}, which consists of a cluster of four gyrotropic spheres. These spheres are assumed to be made of the same materials, but may subject to different local magnetic field $B_i$ (including  those generated by internal magnetization). The dielectric permittivities of the spheres are:
\begin{align}
\label{eq:epsilon_general}
   \varepsilon_{i} = \varepsilon(B_i) = \begin{pmatrix}
    \epsilon_x & -i\epsilon'(B_i) & 0 \\
    i\epsilon'(B_i) & \epsilon_x & 0 \\
    0 & 0 & \epsilon_z
    \end{pmatrix},
\end{align}
where $\epsilon'(B)=-\epsilon'(-B)$ such that $\varepsilon(B)^T = \varepsilon(-B)$ per generalized Onsager reciprocal relations \cite{onsager1931a,onsager1931,casimir1945}. Here for illustration purposes we choose $B_1 = B_3 = -B_2 = - B_4$. For this system, there are two classes of symmetries: (1) the usual point group symmetry $D_{2h} = \{E, C_2(z), 2C_2'(x), i, \sigma_h(z), 2\sigma_v(x) \}$; (2) the compound symmetry $R=  \{2\mathcal{T}C_4(z), 2\mathcal{T}C''_2(a), 2\mathcal{T}S_4(z), 2\mathcal{T}\sigma_d(a)\}$ , where $\mathcal{T}$ is a so-called  “anti-symmetry” operation that transforms ${{\varepsilon}}(\bm{r})$ to ${{\varepsilon^T}}( \bm{r})$, which is equivalent to reversing the direction of the local magnetic field. We use the standard Schoenflies notations~\cite{bradley2010}, and for clarity, we indicate the rotation axis and the normal direction of the mirror plane in the parentheses. \note{For example, $C_2(z)$ denotes  $180^\circ$ rotation around the $z$-axis, $\sigma_v(x)$ denotes the mirror operation with respected to the $y$-$z$ plane, while $\mathcal{T} C_4(z)$ denotes $90^\circ$ rotation around the $z$-axis combined with the  anti-symmetry operation.} We note that each compound symmetry $R_n= \mathcal{T} A_n$ is a combination of $\mathcal{T}$ and a usual spatial operation $A_n$, but $\mathcal{T}$ and $A_n$ are not the symmetry by themselves. All $\{A_n\}$ (not $\{R_n\}$!) together with the usual point group $D_{2h}$ forms a larger point group $D_{4h}$. 
The symmetry group of such a gytropic cluster is therefore a magnetic group:~  
\begin{align}
    \label{eq:Magnetic_group}
    \mathcal{M} =& \underline{4}/mmm = D_{4h}(D_{2h}) \equiv D_{2h} + \mathcal{T}(D_{4h} - D_{2h}) \notag\\
    =&\{E,\, 2\underline{C_4}(z),\, C_2(z),\, 2C'_2(x),\, 2\underline{C''_2}(a),\notag\\
    &\ \ i,\, 2\underline{S_4}(z),\, \sigma_h(z),\, 2\sigma_v(x),\, 2\underline{\sigma_d}(a) \, \},
\end{align}
where the underline denotes the compound elements that are combined with $\mathcal{T}$.
A magnetic group contains the usual point group symmetry operations, as well as compound symmetry operations which contain an “anti-symmetry” operator  $\mathcal{T}$ \cite{bradley2010,hamermesh1989,dresselhaus2010}.

The generalization of the discussion above to an arbitrary system  is straightforward. We define a local operation of anti-symmetry $\mathcal{T}$, which transforms between a general medium as described by ${{C}}(\omega, \bm{r})$ and its complementary medium as described by   ${{\widetilde{C}}}(\omega, \bm{r})$ \cite{jinaukong1972}:
\begin{equation}
\label{eq:T_operation}
    {{C}} = 
    \begin{pmatrix}
    {{\varepsilon}} & {{\zeta}} \\
    {{\eta}} & {{\mu}}
    \end{pmatrix}
    \xleftrightarrow{\enspace \mathcal{T} \enspace}     {{\widetilde{C}}} = 
    \begin{pmatrix}
    {{\varepsilon}}^T & -{{\eta}}^T \\
    -{{\zeta}}^T & {{\mu}}^T 
    \end{pmatrix}.
\end{equation}
$\mathcal{T}^2 = E$, where $E$ is the identity operation. By this definition, for a gyrotropic plasma under an external DC magnetic field bias, its complementary medium is the same gyrotropic plasma but with the direction of the magnetic field bias reversed. A medium is reciprocal if and only if it is self-complementary, i.e.~it is invariant under $\mathcal{T}$. Since $\mathcal{T}$ acts on the constitutive matrix instead of the ordinary position coordinates, it  commutes with all the ordinary spatial operations. The symmetry of a general system  consists of ordinary spatial symmetry operations, and their combination with $\mathcal{T}$. Following Ref.~\cite{bradley2010}, we refer to the former type of operations as uncolored, and the latter type as colored.  The magnetic group of a system as described by ${{C}}(\omega, \bm{r})$ are the sets of all the  symmetry operations that leave the system invariant.

Below we consider the constraints on $\mathcal{F}$ as imposed by the two different classes of symmetry operations: 
\begin{enumerate}
    \item Uncolored operation   $A_l$:
    
    $A_l$ can be represented by the permutation of the bodies:
    \begin{equation}
        \label{eq:permutation_Al}
        \mathbb{P}_{A_l} = \begin{pmatrix}
        0 & 1 & 2 & \ldots & N \\
        0 & P_1 & P_2 & \ldots & P_N
        \end{pmatrix},
    \end{equation}
    We note the environment is invariant under permutation of the bodies ($0\rightarrow 0$). Such a permutation leaves the system invariant, thus it  enforces the  constraints
    \begin{equation}
    \label{eq:constraint_Al}
        F_{P_i\rightarrow P_j} = F_{i\rightarrow j}, \quad i, j = 0, \ldots, N
    \end{equation}
    In matrix form, Eq.~(\ref{eq:constraint_Al}) can be written as  
    \begin{equation}
        \label{eq:constraint_Al_matrix}
        \mathbf{P}_{A_l}\, \mathcal{F} \, \mathbf{P}_{A_l}^T = \mathcal{F},
    \end{equation}
    where $\mathbf{P}_{A_l}$ is the permutation matrix corresponding to $\mathbb{P}_{A_l}$.
    \item Colored operation $R_n = \mathcal{T} A_n$: 
    
    $A_n$ permutates the bodies by \begin{equation}
       \label{eq:permutation_An}
        \mathbb{P}_{A_n} = \begin{pmatrix}
        0 & 1 & 2 & \ldots & N \\
        0 & P'_1 & P'_2 & \ldots & P'_N
        \end{pmatrix},
    \end{equation} 
    The resultant system is complementary to the original one, and an additional $\mathcal{T}$ operation maps the system back to the original one. 
    
    The generalized reciprocity theorem \cite{jinaukong1972} of electromagnetism requires that the exchange  matrices $\mathcal{F}$ and $\widetilde{\mathcal{F}}$  of two  complementary systems   ${{C}}(\omega, \bm{r})$ and   ${{\widetilde{C}}}(\omega, \bm{r})$  are transpose of each other  (see the proof in the Appendix):
    \begin{equation}
        \label{eq:F_transpose}
        \widetilde{\mathcal{F}} = \mathcal{F}^T, \quad i.e.\; \widetilde{F}_{i\rightarrow j} = F_{j\rightarrow i}.
    \end{equation}
Therefore, $R_n$, being a symmetry of the system, enforces the  constraints
    \begin{equation}
    \label{eq:constraint_Rm}
        F_{P'_i\rightarrow P'_j} = F_{j\rightarrow i}, \quad i, j = 0, \ldots, N
    \end{equation}
    In matrix form, Eq.~(\ref{eq:constraint_Rm}) can be written as  
    \begin{equation}
        \label{eq:constraint_Rm_matrix}
        \mathbf{P}_{A_n}\, \mathcal{F} \, \mathbf{P}_{A_n}^T = \mathcal{F}^T,
    \end{equation}
    where $\mathbf{P}_{A_n}$ is the permutation matrix corresponding to $\mathbb{P}_{A_n}$.
\end{enumerate}
By considering all the symmetry elements $\{A_l, R_n\}$ in the magnetic group $\mathcal{M}$, we obtain \textit{all} the constraints imposed by the symmetry on radiative heat transfer.

Magnetic groups,  denoted as $\mathcal{M}$,  can be classified into three types~\cite{bradley2010,dresselhaus2010}:
\begin{enumerate}
    \item Colorless group. Here $\mathcal{M}$ is ordinary point group $\mathcal{G}$ with no colored elements. 
    \item Gray group. Here $\mathcal{M}$ is isomorphic to a direct product $\mathcal{G} \otimes \{E, \mathcal{T}\}$, where $\mathcal{G}$ is a ordinary  point group. Being a direct product immediately implies that $\mathcal{T}$ commutes with all elements of the point group $\mathcal{G}$. 
    
    \item Black/white group. Here $\mathcal{M} = \{A_l, R_n\}$, where half of the elements are colorless forming the set $\{A_l\}$ and the other half are colored forming the set $\{R_n = \mathcal{T} A_n\}$. Moreover, $\{A_l, A_n\}$ forms an ordinary point group $\mathcal{G}'$ and $\{A_l\} = \mathcal{H}$ forms a subgroup of $\mathcal{G'}$ with index 2. Thus a black/white group is  of the form
    \begin{equation}
        \label{eq:black_white}
        \mathcal{M} = \mathcal{H} + \mathcal{T}(\mathcal{G}'-\mathcal{H})
    \end{equation}
    We denote $\mathcal{M} = \mathcal{G}'(\mathcal{H})$ following Ref.~\cite{dresselhaus2010}.
\end{enumerate}
A reciprocal system is by definition invariant under $\mathcal{T}$. Since $\mathcal{T}$ is an element only of a gray group,  a system is reciprocal if and only if its symmetry is a gray group; a system is nonreciprocal if and only its symmetry is a colorless or black/white group.

Finally we consider the constraints of the second law of thermodynamics. Since we consider the bodies exchanging energy only by radiation, in the equilibrium case where all bodies as well as the environment have the same temperature, the energy flow into any body must balances that out of the body:
\begin{equation}
    \label{eq:energy_conservation}
    \sum_{j=0 ;\, j\neq i}^{N} F_{i\rightarrow j} = \sum_{j=0 ;\, j\neq i}^{N} F_{j\rightarrow i}, 
\end{equation}
i.~e.~$\mathcal{F}$ matrix has the same row sum and column sum. This represents a nodal conservation law of heat flow at equilibrium. In matrix form, Eq.~(\ref{eq:energy_conservation}) can be written as 
\begin{equation}
    \label{eq:energy_conservation_matrix}
    (\mathcal{F} - \mathcal{F}^T ) \Vec{j} = 0, 
\end{equation}
where $\Vec{j}$ is an all-one vector. Conversely, if Eq.~(\ref{eq:energy_conservation_matrix}) is satisfied, in equilibrium the net heat flow into any of the subsystem consisting of a few of bodies is zero. Thus Eq.~(\ref{eq:energy_conservation_matrix}) is sufficient to impose the second law of thermodynamics in the many-body system.

\note{The second law of thermodynamics can provide unique constraints beyond  those from  symmetry. For example, a system where radiative heat transfer occurs entirely between two bodies has an exchange matrix
\begin{equation}
    \label{eq:2_2_F_mat}
    \mathcal{F} = 
    \begin{pmatrix}
    0 & F_{1\rightarrow2} \\
    F_{2\rightarrow1} & 0 
    \end{pmatrix}.
\end{equation}
The second law of thermodynamics requires $F_{1\rightarrow2}=F_{2\rightarrow1}$, regardless of any symmetry~\cite{zhu2016}.}

The three sets of constraints Eq.~(\ref{eq:constraint_Al}-\ref{eq:constraint_Al_matrix}), Eq.~(\ref{eq:constraint_Rm}-\ref{eq:constraint_Rm_matrix}) and Eq.~(\ref{eq:energy_conservation}-\ref{eq:energy_conservation_matrix}) are the main results of this paper. These are all the constraints on radiative heat trasfer that can be stated from  symmetry and the second law of thermodynamics. 

Let us apply the general theory  to the concrete example in Fig.~\ref{fig:scheme}. The  magnetic group of the system is  $\mathcal{M} =D_{4h}(D_{2h})$ (Eq.~\ref{eq:Magnetic_group}). 
The exchange matrix is:
\begin{equation}
    \label{eq:5_5_F_mat}
    \mathcal{F} = 
    \begin{pmatrix}
    0 & F_{01} & F_{02} & F_{03} & F_{04} \\
    F_{10} & 0  & F_{12} & F_{13} & F_{14} \\
    F_{20} & F_{21}  & 0 & F_{23} & F_{24} \\
    F_{30} & F_{31}  & F_{32} & 0 & F_{34} \\
    F_{40} & F_{41}  & F_{42} & F_{43} & 0
    \end{pmatrix},
\end{equation}
where $F_{ij}\equiv F_{i\rightarrow j}$ for simplicity.
We first study the constraints on $\mathcal{F}$ imposed 
by $\mathcal{M}$ by considering  all the elements:
\begin{itemize}
    \item 2$\underline{C_4}(z)$: ${C_4}(z)$ permutates the bodies by 
    \begin{equation}
       \label{eq:permutation_C4}
        \mathbb{P}_{C_4} = \begin{pmatrix}
        0 & 1 & 2 & 3 & 4 \\
        0 & 2 & 3 & 4 & 1
        \end{pmatrix},
    \end{equation} 
    Constraints from Eq.~(\ref{eq:constraint_Rm}) are therefore:
    \begin{align}
       \label{eq:constraints_C4}
        F_{01} = F_{20} = F_{03} = F_{40}, \notag\\
        F_{02} = F_{30} = F_{04} = F_{10},\notag\\
        F_{12} = F_{32} = F_{34} = F_{14},\notag\\
        F_{23} = F_{43} = F_{41} = F_{21},\notag\\
        F_{13} = F_{42} = F_{31} = F_{24}.           
    \end{align}
    \item $C_2(z)$: no new constraints, since $C_2(z) = \underline{C_4}^2(z)$. \item $2C_2'(x)$: permutates the bodies by: 
    \begin{equation}
       \label{eq:permutation_C2'}
        \mathbb{P}_{C'_2} = \begin{pmatrix}
        0 & 1 & 2 & 3 & 4 \\
        0 & 4 & 3 & 2 & 1
        \end{pmatrix},
    \end{equation} 
    Constraints from Eq.~(\ref{eq:constraint_Al}) are therefore:
    \begin{align}
       \label{eq:constraints_C2'}
        F_{01} = F_{04} , \quad F_{12} = F_{43}.
    \end{align}
    \item The remaining elements impose no new constraints.
\end{itemize}
Combining all the constraints Eq.~(\ref{eq:constraints_C4}) and~(\ref{eq:constraints_C2'}), 
\begin{equation}
    \label{eq:5_5_F_mat_constrained}
    \mathcal{F} = 
    \begin{pmatrix}
    0 & F_{01} & F_{01} & F_{01} & F_{01} \\
    F_{01} & 0  & F_{12} & F_{13} & F_{12} \\
    F_{01} & F_{12}  & 0 & F_{12} & F_{13} \\
    F_{01} & F_{13}  & F_{12} & 0 & F_{12} \\
    F_{01} & F_{12}  & F_{13} & F_{12} & 0
    \end{pmatrix},
\end{equation}
which has only 3 independent components $F_{01}, F_{12}, F_{13}$. Also $\mathcal{F} = \mathcal{F}^T$, even though this system is nonreciprocal.

The second law of thermodynamics imposes no new constraints, since $\mathcal{F} = \mathcal{F}^T$, and Eq.~(\ref{eq:energy_conservation_matrix}) is automatically satisfied.

\section{Applications}
\label{sec:application}
As an application of our theory, we study the persistent heat current in nonreciprocal radiative heat transfer. Persistent heat current is a  phenomenon that can exist in some nonreciprocal many-body systems even at thermal equilibrium \cite{zhu2016}. By definition, the persistent heat current exists between body $i$ and $j$ at equilibrium if and only if $F_{i\rightarrow j} \neq F_{j\rightarrow i}$. It has been proved that nonreciprocity is a necessary but not sufficient condition for the existence of persistent heat current~\cite{zhu2016,guo2019a}. 
However,  there still lacks a systematic way to determine whether a given system can exhibit persistent heat current. Our theory can provide  such a systematic approach. 

From the definition, there is persistent heat current in a system between at least one pair of bodies, if and only if  $\mathcal{F} \neq \mathcal{F}^T$. Since our theory provides all the general constraints on $\mathcal{F}$, we can check  whether a system can support persistent heat current by deducing the constrained form of  $\mathcal{F}$ and then checking whether $\mathcal{F}=\mathcal{F}^T$. If $\mathcal{F}=\mathcal{F}^T$, there is no persistent heat current in such a system. Otherwise, there is no symmetry reason against the existence of persistent heat current.
\begin{figure}[htbp]
    \centering
    \includegraphics[width=0.95\columnwidth]{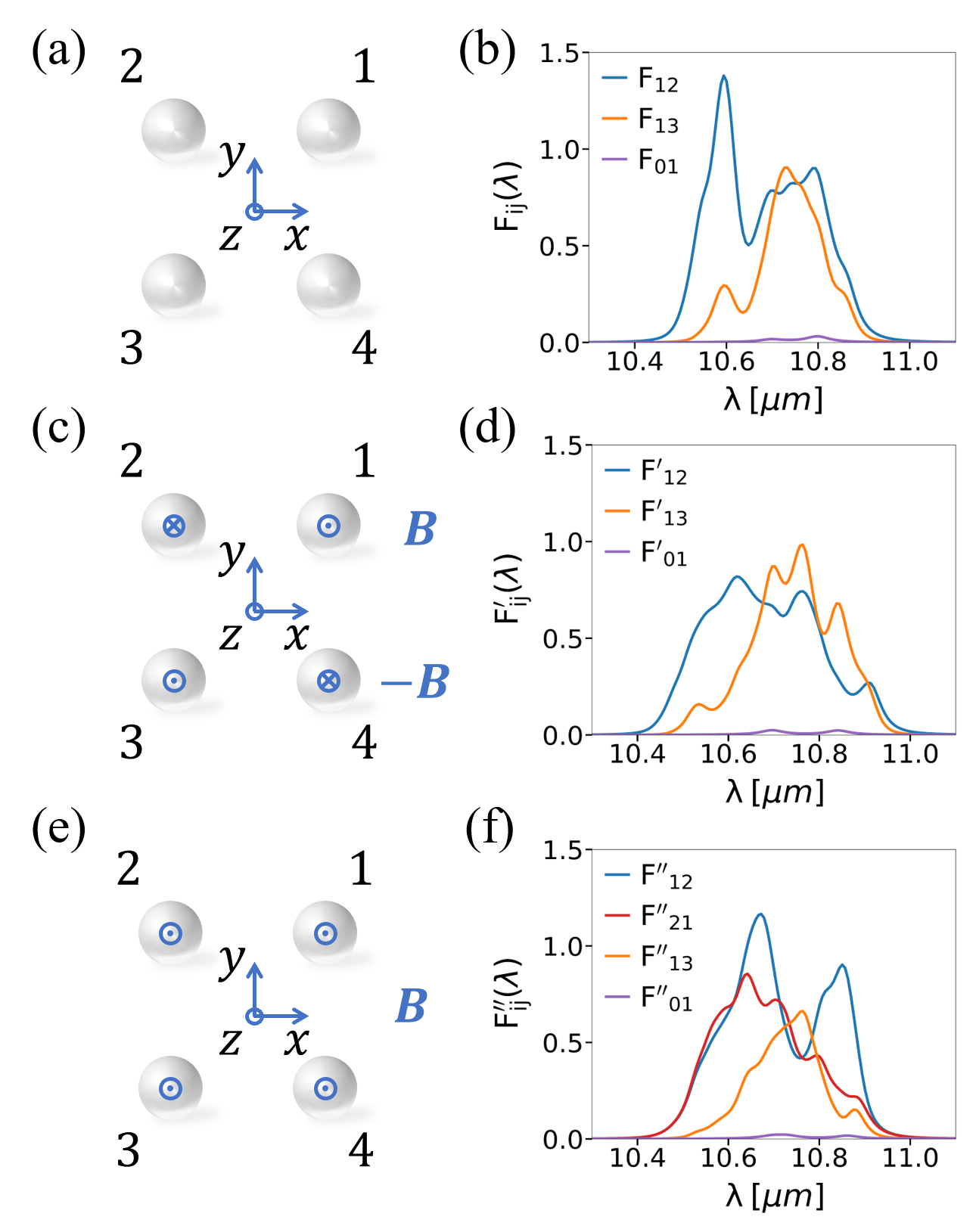}
    \caption{(a) A reciprocal system that exhibits no persistent heat current. It consists of four InSb  spheres with centers placed at the vertices of a square on the $x$-$y$  plane under no external magnetic field.  (b) The calculated transmission coefficient spectra $F_{ij}(\lambda)$ for the system in (a). Only the independent components are plotted. (c) A nonreciprocal system that exhibits no  persistent heat current. It consists of the same spheres as (a), but under an  external magnetic field along the $z$ direction with alternating strength in the $x$-$y$ plane: $B_1 = B_3 = -B_2 = -B_4 = 1$ T. (d) The calculated independent  transmission coefficient spectra $F'_{ij}(\lambda)$ for the system in (c). (e) A nonreciprocal system that exhibits   persistent heat current. It consists of the same spheres as (a), but under a uniform external magnetic field along the $z$ direction: $B_1 = B_2 = B_3 = B_4 = 1$ T. (f) The calculated independent  transmission coefficient spectra $F''_{ij}(\lambda)$ for the system in (e).}
    \label{fig:spheres}
\end{figure}

To demonstrate such a procedure, we consider three exemplary systems as shown in Fig.~\ref{fig:spheres}(a,c,e). The geometries are similar: all the systems consist of  four gyrotropic spheres made of $n$-doped InSb. Each sphere has a radius of $100$ nm. The centers of the four spheres are placed at the vertices of a square on the $x$-$y$  plane. The side length of the square is $320$ nm. These systems differ in the magnetic field configurations: the first system is under no magnetic field, the second under spatially alternating fields ($B_1 = B_3 = -B_2 = -B_4 = 1$ T), and the last under a uniform field ($B_1 = B_2 = B_3 = B_4 = 1$ T). The external magnetic field is perpendicular to the $x$-$y$  plane. Under the magnetic field, $n$-doped InSb has a relative permittivity tensor
\begin{equation*}
    \overline{\overline{\epsilon}} = \epsilon_b \overline{\overline{I}} - \frac{\omega_p^2}{(\omega+i\Gamma)^2-\omega_c^2} 
    \begin{bmatrix}
    1 + i\frac{\Gamma}{\omega} & -i\frac{\omega_c}{\omega} & 0 \\
    i\frac{\omega_c}{\omega} & 1 + i\frac{\Gamma}{\omega} & 0 \\
    0 & 0 & \frac{(\omega+i\Gamma)^2-\omega_c^2}{\omega (\omega+i\Gamma)}
    \end{bmatrix}.
\end{equation*}
Here, the first term is the background permittivity as taken from Ref.~\cite{palik1998handbook}. The second term takes into account free-carrier contribution, which is sensitive to external magnetic field. $\Gamma$ is the free-carrier relaxation rate, $\omega_c = eB/m^*$ is the cyclotron frequency, and $\omega_p = \sqrt{n_e e^2/(m^*\epsilon_0)}$
is the plasma frequency. For calculation, we use $n_e =1.36 \times 10^{19}\, \mathrm{cm^{-3}}$,   
 $ \Gamma = 10^{12}\, \mathrm{s^{-1}}$ and  $m^* = 0.08\,m_e$. 

The three systems have  different symmetries. The first system is reciprocal and  has a gray group $\mathcal{M}_1 = D_{4h} \otimes \{E, \mathcal{T}\}$. The second system, which is identical to that in Fig.~\ref{fig:scheme},  has a black/white group $\mathcal{M}_2 = D_{4h}(D_{2h})$. The last system has a black/white group $\mathcal{M}_3 = D_{4h}(C_{4h})$. Consequently, the three constrained $\mathcal{F}$ matrices have the following forms respectively:    
\begin{equation}
    \label{eq:F1_mat_constrained}
    \mathcal{F}_1 = 
    \begin{pmatrix}
    0 & F_{01} & F_{01} & F_{01} & F_{01} \\
    F_{01} & 0  & F_{12} & F_{13} & F_{12} \\
    F_{01} & F_{12}  & 0 & F_{12} & F_{13} \\
    F_{01} & F_{13}  & F_{12} & 0 & F_{12} \\
    F_{01} & F_{12}  & F_{13} & F_{12} & 0
    \end{pmatrix},
\end{equation}
\begin{equation}
    \label{eq:F2_mat_constrained}
    \mathcal{F}_2 = 
    \begin{pmatrix}
    0 & F'_{01} & F'_{01} & F'_{01} & F'_{01} \\
    F'_{01} & 0  & F'_{12} & F'_{13} & F'_{12} \\
    F'_{01} & F'_{12}  & 0 & F'_{12} & F'_{13} \\
    F'_{01} & F'_{13}  & F'_{12} & 0 & F'_{12} \\
    F'_{01} & F'_{12}  & F'_{13} & F'_{12} & 0
    \end{pmatrix},
\end{equation}
\begin{equation}
    \label{eq:F3_mat_constrained}
    \mathcal{F}_3 = 
    \begin{pmatrix}
    0 & F''_{01} & F''_{01} & F''_{01} & F''_{01} \\
    F''_{01} & 0  & F''_{12} & F''_{13} & F''_{21} \\
    F''_{01} & F''_{21}  & 0 & F''_{12} & F''_{13} \\
    F''_{01} & F''_{13}  & F''_{21} & 0 & F''_{12} \\
    F''_{01} & F''_{12}  & F''_{13} & F''_{21} & 0
    \end{pmatrix}.
\end{equation}

We make several  observations. The first system in Fig.~\ref{fig:spheres}(a) is reciprocal, thus cannot exhibit persistent heat current  as expected ($\mathcal{F}_1 = \mathcal{F}_1^T$). Moreover, $\mathcal{F}_1$ has only 3 independent components $F_{01}$, $F_{12}$ and $F_{13}$ as required by symmetry. The second system in Fig.~\ref{fig:spheres}(c), even though is nonreciprocal, cannot exhibit persistent heat current either since $\mathcal{F}_2 = \mathcal{F}_2^T$. Interestingly, $\mathcal{F}_2$ has exactly the same form as $\mathcal{F}_1$. This highlights the possibility  that many-body systems with completely different symmetries can exhibit the same qualitative behavior in radiative heat transfer. The last system in Fig.~\ref{fig:spheres}(e) is  nonreciprocal and can hold persistent heat current as $\mathcal{F}_3 \neq \mathcal{F}_3^T$. $\mathcal{F}_3$ has 4   independent components $F''_{01}$, $F''_{12}$, $F''_{21}$ and $F''_{13}$.    

We numerically verify these observations by calculating  the transmission coefficient spectra $F_{ij}(\lambda)$ in Figs.~\ref{fig:spheres}(b,d,f) corresponding to  the systems in Figs.~\ref{fig:spheres}(a,c,e), respectively. We plot all the independent components of the $\mathcal{F}$ matrices, and verify that
the other components indeed obey the relations in Eq.~(\ref{eq:F1_mat_constrained}-\ref{eq:F3_mat_constrained}). We see there is no persistent heat (e.g.~$F_{12} = F_{21}$ and  $F'_{12} = F'_{21}$) in the first and the second systems, while there is persistent heat current in the last system ($F''_{12} \neq F''_{21}$). 

As another application of our theory, we have the following proposition:
if a system has a colored symmetry $R = \mathcal{T} A$
where $\mathbb{P}_A = \mathbb{I}$ is an identity permutation, it cannot exhibit persistent heat current. This is because if there is such an element $R$, its constraint on $\mathcal{F}$ (Eq.~(\ref{eq:constraint_Rm_matrix})) is:
\begin{equation}
    \label{eq:proof_R}
    \mathbf{I} \mathcal{F} \mathbf{I} = \mathcal{F}^T,
\end{equation}
where $\mathbf{I}$ is the identity matrix. Therefore $\mathcal{F} = \mathcal{F}^T$, which precludes persistent heat current.

\begin{figure}[htbp]
    \centering
    \includegraphics[width=1.0\columnwidth]{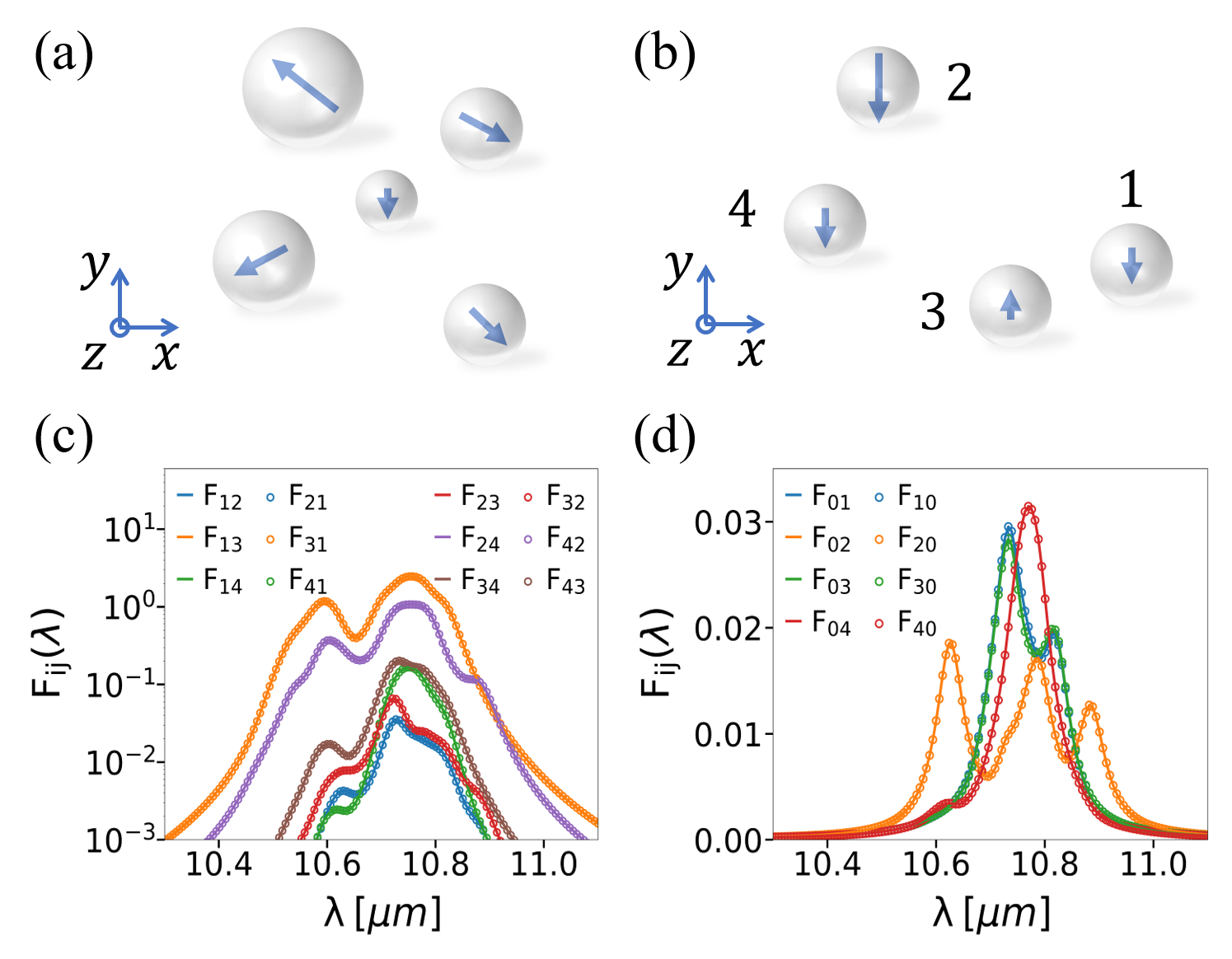}
    \caption{(a) A cluster of magnetooptical  spheres with their centers placed on $x$-$y$ plane under an inhomogeneous external  magnetic field parallel to $x$-$y$ plane. Such a system exhibits no persistent heat current. (b) A cluster of four InSb spheres with their centers placed on $x$-$y$ plane. The spheres are of the same size with a radius of $100$ nm. The spheres experience different  magnetic fields along the $y$ direction: $B_1 = -0.256$ T, $B_2 = -1.896$ T, $B_3 = 0.199$ T, $B_4 = -0.259$ T. (c) The  transmission coefficient spectra   
   between the spheres. The logarithmic scale is used to accommodate the different magnitudes.  (d) The transmission coefficient spectra  between the spheres and the environment. (c,d) shows $F_{ij}(\lambda) = F_{ji}(\lambda)$ thus there is no persistent heat current. }
    \label{fig:random}
\end{figure}

As an example, we consider clusters of gyrotropic  spheres with their centers lying on a plane under an external magnetic field parallel to that plane. The spheres can be of different sizes, and the magnetic field can be inhomogeneous.  One typical  system is depicted in Fig.~\ref{fig:random}(a). Such systems have the $\mathcal{T} m$ symmetry, where $m$ is the mirror operation with respect to the plane passing through the  centers, and $\mathbb{P}_m = \mathbb{I}$. Therefore, such systems cannot exhibit persistent heat current. \note{In contrast, clusters of randomly positioned gyrotropic particles subjected to magnetic field with random magnitudes and directions in general does not have $\mathcal{T} m$ symmetry and can therefore exhibit persistent heat current.}

We now provide numerical evidences. For the convenience of simulation,  instead of  Fig.~\ref{fig:random}(a), we consider the system in Fig.~\ref{fig:random}(b), where the four InSb spheres are of the same size with a radius of $100$ nm, and the magnetic field is inhomogeneous but along the same ($y$)  direction. The centers of the spheres are placed on the $x$-$y$ plane 
with randomly chosen coordinates (unit: nm): $(484, -146),\, (-167,313),\, (174, -252),\,$ and $(-303,-41)$. The spheres are  under randomly assigned magnetic fields $B_1 = -0.256$ T, $B_2 = -1.896$ T, $B_3 = 0.199$ T, and $B_4 = -0.259$ T. Such a system has a black/white magnetic group $C_{1h}(C_1) = \{E, \mathcal{T}m\}$. The only constraints that can be deduced are  $F_{ij} = F_{ji}$. Thus  there are $10$ independent components for the $\mathcal{F}$ matrix. We numerically calculate the transmission coefficient spectra $F_{ij}(\lambda)$. Fig.~\ref{fig:random}(c) plots the transmission coefficients between  bodies. Here we use logarithmic scale to accommodate the different magnitudes.  Fig.~\ref{fig:random}(d) plots the transmission coefficients between the  bodies and the environment. These two plots confirm that $F_{ij}(\lambda) = F_{ji}(\lambda)$, $0\leq i<j \leq 4$, and there are indeed $10$ independent components.

\section{Discussion and Conclusion}
\label{sec:conclusion}
\note{Throughout the paper, we have used clusters of spherical particles subjected to local magnetic field as concrete examples to illustrate the theory. Our theory, which is based on symmetry argument alone, is not restricted to either spherical particles or sub-wavelength particles, but is applicable to arbitrary many-body systems, which can include non-spherical objects, or objects with sizes comparable or larger than the relevant thermal wavelengths.}

In conclusion, we have studied the constraints on many-body radiative heat transfer imposed by symmetry  and the second law of thermodynamics. We show that the symmetry of these systems in general can be described by a magnetic group. And the constraints of the magnetic group on heat transfer can be derived using the generalized reciprocity theorem.  
We also show that the the second law of thermodynamics provides additional  constraints in the form of a nodal conservation law of heat flow at equilibrium. As an  application of the theory, we provide a systematic approach to determine the existence of persistent heat current in arbitrary many-body systems. Our work should be useful in providing theoretical guidance for exploring novel effects of radiative heat transfer in complex many-body systems and networks.  

\appendix*
\section{Proof of Eq.~(\ref{eq:F_transpose})}
We prove Eq.~(\ref{eq:F_transpose}) using the  generalized reciprocity theorem \cite{jinaukong1972}. First we briefly review Lorentz reciprocity~\cite{rumsey1954}. Consider two sources $\bm{J_a}$ and $\bm{J_b}$, which produce fields $\bm{E}_a$ and $\bm{E}_b$, respectively. The Lorentz reciprocity theorem states that 
for a reciprocal medium that satisfies  ${{C}} = {{\widetilde{C}}}$,
\begin{equation}
    \label{eq:oridnary_reciprocity}
    \iiint_V dV \bm{E}_a\cdot \bm{J}_b =     \iiint_V dV \bm{E}_b\cdot \bm{J}_a,
\end{equation}
where the integration is  over the volume that contains the sources $a$ and $b$. 

The reciprocity theorem stated above can be generalized to arbitrary media  \cite{villeneuve1958,jinaukong1972}. Consider two sources $\bm{J_a}$ and $\bm{J_b}$, producing fields $\bm{E}_a$ and $\bm{E}_b$ in the original medium ${{C}}$, and fields $\bm{\widetilde{E}}_a$ and $\bm{\widetilde{E}}_b$ in the complementary medium ${{\widetilde{C}}}$, respectively. Then the generalized  reciprocity theorem states
\begin{equation}
    \label{eq:general_reciprocity}
    \iiint_V dV \bm{E}_a\cdot \bm{J}_b =     \iiint_V dV \bm{\widetilde{E}}_b\cdot \bm{J}_a, 
\end{equation}
integrating over the volume that contains the sources $a$ and $b$.
Note it reduces to the ordinary reciprocity theorem for reciprocal medium.

The generalized reciprocity theorem  requires the dyadic Green's functions for the corresponding bodies in the original and complementary systems to be transpose of each other:
\begin{equation}
    \label{eq:green_function_transpose}
    \widetilde{\mathbb{G}}_\alpha(\bm{r},\bm{r}') = \mathbb{G}^T_\alpha(\bm{r}',\bm{r}),   
\end{equation}
where body $\alpha$ can be a composite consisting of multiple bodies. 
 
Consequently, the $\mathbb{T}$ operators for the corresponding bodies in the two systems are also transpose of each other:
\begin{equation}
    \label{eq:T_matrix_transpose}
    \widetilde{\mathbb{T}}_\alpha (\bm{r},\bm{r}')= \mathbb{T}^T_\alpha (\bm{r}',\bm{r}).
\end{equation}
This follows from the definition of  $\mathbb{T}$ \cite{kruger2012,zhu2018},
\begin{align}
    \label{eq:G_T}
    \mathbb{G}_\alpha &= \mathbb{G}_0 + \mathbb{G}_0 \mathbb{T}_\alpha \mathbb{G}_0,\\
    \widetilde{\mathbb{G}}_\alpha &= \mathbb{G}_0 + \mathbb{G}_0 \widetilde{\mathbb{T}}_\alpha \mathbb{G}_0,
\end{align}
where $\mathbb{G}_0$, being  the free-space Green's function, is symmetric, i.~e.~$\mathbb{G}_0(\bm{r}, \bm{r}') = \mathbb{G}_0^T (\bm{r}',\bm{r})$.

Now we proceed to prove that the exchange matrices $\mathcal{F}$ and $\mathcal{F'}$ of the original and complementary systems are transpose of each other (Eq.~(\ref{eq:F_transpose})). Without loss of generality, we consider heat exchange between bodies 1 and 2 in the original many-body system, while body $3$ includes all other bodies except $1,2$. We label the corresponding bodies in the complementary system as $1',2',3'$. In the derivation below, we follow a similar procedure in the supplement of Ref.~\cite{zhu2016}, and suppress the parameters of the operators for clarity.

The spectral transmission coefficient for heat transfer to body 2 due to thermal noise of body 1 is:
\begin{equation}
    \label{eq:F_12}
    F_{1\rightarrow2}(\omega) = 4 \operatorname{Tr}[\mathbb{Q}_2 \mathbb{W}_{21} \mathbb{R}_1 \mathbb{W}_{21}^\dagger],
\end{equation}
while that to body $2'$ due to thermal noise of body $1'$ is:
\begin{align}
    \label{eq:F'_21}
    \widetilde{F}_{2'\rightarrow1'}(\omega) &= 4 \operatorname{Tr}[\widetilde{\mathbb{Q}}_{1} \widetilde{\mathbb{W}}_{12} \widetilde{\mathbb{R}}_2 \widetilde{\mathbb{W}}_{12}^\dagger] \notag \\
    &= 4 \operatorname{Tr}[\widetilde{\mathbb{W}}_{12}^*\widetilde{\mathbb{R}}_{2}^T \widetilde{\mathbb{W}}_{12}^T \widetilde{\mathbb{Q}}_1^T ] \notag \\
    &= 4 \operatorname{Tr}[\widetilde{\mathbb{R}}_{2}^T \widetilde{\mathbb{W}}_{12}^T \widetilde{\mathbb{Q}}_1^T \widetilde{\mathbb{W}}_{12}^*],
\end{align}
where we have performed transposition of the matrix  product in the second line,  and cyclic permutation in the third line. Here,
\begin{align}
    \label{eq:R}
    \mathbb{R}_\alpha &= \mathbb{G}_0 [\frac{\mathbb{T}_\alpha-\mathbb{T}_\alpha^\dagger}{2i} - \mathbb{T}_\alpha \operatorname{Im}(\mathbb{G}_0) \mathbb{T}_\alpha^\dagger] \mathbb{G}_0^\dagger , \\
    \widetilde{\mathbb{R}}_\alpha^T &= \{\mathbb{G}_0 [\frac{\widetilde{\mathbb{T}}_\alpha-\widetilde{\mathbb{T}}_\alpha^\dagger}{2i} - \widetilde{\mathbb{T}}_\alpha \operatorname{Im}(\mathbb{G}_0) \widetilde{\mathbb{T}}_\alpha^\dagger] \mathbb{G}_0^\dagger\}^T \notag\\
    &=\mathbb{G}_0^\dagger [\frac{\mathbb{T}_\alpha-\mathbb{T}_\alpha^\dagger}{2i} - \mathbb{T}_\alpha^\dagger \operatorname{Im}(\mathbb{G}_0) \mathbb{T}_\alpha] \mathbb{G}_0, \label{eq:R_tilde_T}
\end{align}
\begin{align}
    \mathbb{Q}_\alpha &=\mathbb{G}_0^\dagger [\frac{\mathbb{T}_\alpha-\mathbb{T}_\alpha^\dagger}{2i} - \mathbb{T}_\alpha^\dagger \operatorname{Im}(\mathbb{G}_0) \mathbb{T}_\alpha] \mathbb{G}_0, \\
    \widetilde{\mathbb{Q}}_\alpha^T &= \{\mathbb{G}_0^\dagger [\frac{\widetilde{\mathbb{T}}_\alpha-\widetilde{\mathbb{T}}_\alpha^\dagger}{2i} - \widetilde{\mathbb{T}}_\alpha^\dagger \operatorname{Im}(\mathbb{G}_0) \widetilde{\mathbb{T}}_\alpha] \mathbb{G}_0\}^T \notag\\
    &=\mathbb{G}_0 [\frac{\mathbb{T}_\alpha-\mathbb{T}_\alpha^\dagger}{2i} - \mathbb{T}_\alpha \operatorname{Im}(\mathbb{G}_0) \mathbb{T}_\alpha^\dagger] \mathbb{G}_0^\dagger, \label{eq:Q_tilde_T}
\end{align}
where $\alpha = 1, 2$,  and we have used Eq.~(\ref{eq:T_matrix_transpose}) to simplify Eq.~(\ref{eq:R_tilde_T}, \ref{eq:Q_tilde_T}). Therefore,
\begin{equation}
    \label{eq:relation_QR}
    \widetilde{\mathbb{R}}_\alpha^T = \mathbb{Q}_\alpha, \quad \widetilde{\mathbb{Q}}_\alpha^T = \mathbb{R}_\alpha
\end{equation}
And
\begin{widetext}
\begin{align}
    \label{eq:W12}
    \mathbb{W}_{21} &= \mathbb{G}_0^{-1}\frac{1}{1-\mathbb{G}_0\mathbb{T}_3\mathbb{G}_0\mathbb{T}_2}(1+\mathbb{G}_0\mathbb{T}_3)\frac{1}{1-\mathbb{G}_0\mathbb{T}_1[(1+\mathbb{G}_0\mathbb{T}_2)\frac{1}{1-\mathbb{G}_0\mathbb{T}_3\mathbb{G}_0\mathbb{T}_2}(1+\mathbb{G}_0 \mathbb{T}_3)-1]}, \\
    \widetilde{\mathbb{W}}_{12} &= \mathbb{G}_0^{-1}\frac{1}{1-\mathbb{G}_0\widetilde{\mathbb{T}}_3\mathbb{G}_0\widetilde{\mathbb{T}}_1}(1+\mathbb{G}_0 \widetilde{\mathbb{T}}_3)\frac{1}{1-\mathbb{G}_0\widetilde{\mathbb{T}}_2[(1+\mathbb{G}_0\widetilde{\mathbb{T}}_1)\frac{1}{1-\mathbb{G}_0\widetilde{\mathbb{T}}_3\mathbb{G}_0\widetilde{\mathbb{T}}_1}(1+\mathbb{G}_0 \widetilde{\mathbb{T}}_3)-1]} \notag\\
    &= \mathbb{G}_0^{-1} \frac{1}{1-[(1+\mathbb{G}_0\widetilde{\mathbb{T}}_3)\frac{1}{1-\mathbb{G}_0\widetilde{\mathbb{T}}_2\mathbb{G}_0\widetilde{\mathbb{T}}_3}(1+\mathbb{G}_0\widetilde{\mathbb{T}}_2)-1]\mathbb{G}_0\widetilde{\mathbb{T}}_1}(1+\mathbb{G}_0 \widetilde{\mathbb{T}}_3)\frac{1}{1-\mathbb{G}_0\widetilde{\mathbb{T}}_2\mathbb{G}_0\widetilde{\mathbb{T}}_3} \notag\\
    &= \frac{1}{1-[(1+\widetilde{\mathbb{T}}_3\mathbb{G}_0)\frac{1}{1-\widetilde{\mathbb{T}}_2\mathbb{G}_0\widetilde{\mathbb{T}}_3\mathbb{G}_0}(1+\widetilde{\mathbb{T}}_2\mathbb{G}_0)-1]\widetilde{\mathbb{T}}_1\mathbb{G}_0}(1+ \widetilde{\mathbb{T}}_3\mathbb{G}_0)\frac{1}{1-\widetilde{\mathbb{T}}_2\mathbb{G}_0\widetilde{\mathbb{T}}_3\mathbb{G}_0} \mathbb{G}_0^{-1}, \label{eq:W12_tilde}\\
     \widetilde{\mathbb{W}}_{12}^T &= \mathbb{G}_0^{-1}\frac{1}{1-\mathbb{G}_0\mathbb{T}_3\mathbb{G}_0\mathbb{T}_2}(1+\mathbb{G}_0\mathbb{T}_3)\frac{1}{1-\mathbb{G}_0\mathbb{T}_1[(1+\mathbb{G}_0\mathbb{T}_2)\frac{1}{1-\mathbb{G}_0\mathbb{T}_3\mathbb{G}_0\mathbb{T}_2}(1+\mathbb{G}_0 \mathbb{T}_3)-1]},    \label{eq:W12_tilde_T}
\end{align}
where in Eq.~(\ref{eq:W12_tilde}) we have used Eq.~(11) in the Supplement of \cite{zhu2016} to get the second line, and rearranged the terms to get the third line. We transpose Eq.~(\ref{eq:W12_tilde}) to obtain Eq.~(\ref{eq:W12_tilde_T}).
Therefore,
\begin{align}
    \label{eq:relation_W}
    \widetilde{\mathbb{W}}_{12}^T = \mathbb{W}_{21}, \quad \widetilde{\mathbb{W}}_{12}^* = \mathbb{W}_{21}^\dagger.
\end{align}
\end{widetext}
Using Eq.~(\ref{eq:relation_QR}) and Eq.~(\ref{eq:relation_W}), Eq.~(\ref{eq:F'_21}) becomes
\begin{align}
    \label{eq:F'_F}
    \widetilde{F}_{2'\rightarrow1'}(\omega) &= 4 \operatorname{Tr}[\widetilde{\mathbb{R}}_{2}^T \widetilde{\mathbb{W}}_{12}^T \widetilde{\mathbb{Q}}_1^T \widetilde{\mathbb{W}}_{12}^*] \notag\\
    &= 4 \operatorname{Tr}[\mathbb{Q}_{2} \mathbb{W}_{21} \mathbb{R}_1 \mathbb{W}_{21}^\dagger] 
\end{align}
Comparing Eq.~(\ref{eq:F'_F}) and Eq.~(\ref{eq:F_12}), we get
\begin{equation}
    \label{eq:result}
    \widetilde{F}_{2'\rightarrow1'}(\omega) = F_{1\rightarrow 2}(\omega)
\end{equation}
Since bodies $1,2$ are    arbitrarily chosen, we have proved
\begin{equation}
    \label{eq:result_matrix}
    \widetilde{\mathcal{F}} = \mathcal{F}^T.
\end{equation}

\begin{acknowledgments}
C. Guo thanks  Dr.~Yu Guo, Dr.~Bo Zhao and Dr.~Linxiao Zhu for helpful discussion. This work is supported by U.~S. Army Research Office (ARO) MURI Grant No.~W911NF-19-1-0279.
\end{acknowledgments}


\begin{thebibliography}{35}%
\makeatletter
\providecommand \@ifxundefined [1]{%
 \@ifx{#1\undefined}
}%
\providecommand \@ifnum [1]{%
 \ifnum #1\expandafter \@firstoftwo
 \else \expandafter \@secondoftwo
 \fi
}%
\providecommand \@ifx [1]{%
 \ifx #1\expandafter \@firstoftwo
 \else \expandafter \@secondoftwo
 \fi
}%
\providecommand \natexlab [1]{#1}%
\providecommand \enquote  [1]{``#1''}%
\providecommand \bibnamefont  [1]{#1}%
\providecommand \bibfnamefont [1]{#1}%
\providecommand \citenamefont [1]{#1}%
\providecommand \href@noop [0]{\@secondoftwo}%
\providecommand \href [0]{\begingroup \@sanitize@url \@href}%
\providecommand \@href[1]{\@@startlink{#1}\@@href}%
\providecommand \@@href[1]{\endgroup#1\@@endlink}%
\providecommand \@sanitize@url [0]{\catcode `\\12\catcode `\$12\catcode
  `\&12\catcode `\#12\catcode `\^12\catcode `\_12\catcode `\%12\relax}%
\providecommand \@@startlink[1]{}%
\providecommand \@@endlink[0]{}%
\providecommand \url  [0]{\begingroup\@sanitize@url \@url }%
\providecommand \@url [1]{\endgroup\@href {#1}{\urlprefix }}%
\providecommand \urlprefix  [0]{URL }%
\providecommand \Eprint [0]{\href }%
\providecommand \doibase [0]{https://doi.org/}%
\providecommand \selectlanguage [0]{\@gobble}%
\providecommand \bibinfo  [0]{\@secondoftwo}%
\providecommand \bibfield  [0]{\@secondoftwo}%
\providecommand \translation [1]{[#1]}%
\providecommand \BibitemOpen [0]{}%
\providecommand \bibitemStop [0]{}%
\providecommand \bibitemNoStop [0]{.\EOS\space}%
\providecommand \EOS [0]{\spacefactor3000\relax}%
\providecommand \BibitemShut  [1]{\csname bibitem#1\endcsname}%
\let\auto@bib@innerbib\@empty
\bibitem [{\citenamefont {Planck}(1991)}]{planck1991}%
  \BibitemOpen
  \bibfield  {author} {\bibinfo {author} {\bibfnamefont {M.}~\bibnamefont
  {Planck}},\ }\href@noop {} {\emph {\bibinfo {title} {The Theory of Heat
  Radiation}}}\ (\bibinfo  {publisher} {{Dover Publications}},\ \bibinfo
  {address} {{New York}},\ \bibinfo {year} {1991})\BibitemShut {NoStop}%
\bibitem [{\citenamefont {Rytov}\ \emph {et~al.}(1989)\citenamefont {Rytov},
  \citenamefont {Kravtsov},\ and\ \citenamefont {Tatarskii}}]{rytov1989}%
  \BibitemOpen
  \bibfield  {author} {\bibinfo {author} {\bibfnamefont {S.~M.}\ \bibnamefont
  {Rytov}}, \bibinfo {author} {\bibfnamefont {Y.~A.}\ \bibnamefont
  {Kravtsov}},\ and\ \bibinfo {author} {\bibfnamefont {V.~I.}\ \bibnamefont
  {Tatarskii}},\ }\href@noop {} {\emph {\bibinfo {title} {Principles of
  Statistical Radiophysics 3: Elements of Random Fields.}}}\ (\bibinfo
  {publisher} {{Springer-Verlag}},\ \bibinfo {address} {{Berlin, Heidelberg}},\
  \bibinfo {year} {1989})\BibitemShut {NoStop}%
\bibitem [{\citenamefont {Chen}(2005)}]{chen2005}%
  \BibitemOpen
  \bibfield  {author} {\bibinfo {author} {\bibfnamefont {G.}~\bibnamefont
  {Chen}},\ }\href@noop {} {\emph {\bibinfo {title} {Nanoscale Energy Transport
  and Conversion: A Parallel Treatment of Electrons, Molecules, Phonons, and
  Photons}}}\ (\bibinfo  {publisher} {{Oxford University Press}},\ \bibinfo
  {address} {{Oxford}},\ \bibinfo {year} {2005})\BibitemShut {NoStop}%
\bibitem [{\citenamefont {Zhang}(2007)}]{zhang2007}%
  \BibitemOpen
  \bibfield  {author} {\bibinfo {author} {\bibfnamefont {Z.~M.}\ \bibnamefont
  {Zhang}},\ }\href@noop {} {\emph {\bibinfo {title} {Nano/Microscale Heat
  Transfer}}}\ (\bibinfo  {publisher} {{McGraw-Hill}},\ \bibinfo {address}
  {{New York}},\ \bibinfo {year} {2007})\BibitemShut {NoStop}%
\bibitem [{\citenamefont {Howell}\ \emph {et~al.}(2016)\citenamefont {Howell},
  \citenamefont {Meng{\"u}{\c c}},\ and\ \citenamefont {Siegel}}]{howell2016}%
  \BibitemOpen
  \bibfield  {author} {\bibinfo {author} {\bibfnamefont {J.~R.}\ \bibnamefont
  {Howell}}, \bibinfo {author} {\bibfnamefont {M.~P.}\ \bibnamefont
  {Meng{\"u}{\c c}}},\ and\ \bibinfo {author} {\bibfnamefont {R.}~\bibnamefont
  {Siegel}},\ }\href@noop {} {\emph {\bibinfo {title} {Thermal Radiation Heat
  Transfer}}},\ \bibinfo {edition} {sixth}\ ed.\ (\bibinfo  {publisher} {{CRC
  Press}},\ \bibinfo {address} {{London}},\ \bibinfo {year} {2016})\BibitemShut
  {NoStop}%
\bibitem [{\citenamefont {Fan}(2017)}]{fan2017}%
  \BibitemOpen
  \bibfield  {author} {\bibinfo {author} {\bibfnamefont {S.}~\bibnamefont
  {Fan}},\ }\bibfield  {title} {\bibinfo {title} {Thermal {{Photonics}} and
  {{Energy Applications}}},\ }\href
  {https://doi.org/10.1016/j.joule.2017.07.012} {\bibfield  {journal} {\bibinfo
   {journal} {Joule}\ }\textbf {\bibinfo {volume} {1}},\ \bibinfo {pages} {264}
  (\bibinfo {year} {2017})}\BibitemShut {NoStop}%
\bibitem [{\citenamefont {Onsager}(1931{\natexlab{a}})}]{onsager1931a}%
  \BibitemOpen
  \bibfield  {author} {\bibinfo {author} {\bibfnamefont {L.}~\bibnamefont
  {Onsager}},\ }\bibfield  {title} {\bibinfo {title} {Reciprocal {{Relations}}
  in {{Irreversible Processes}}. {{I}}.},\ }\href
  {https://doi.org/10.1103/PhysRev.37.405} {\bibfield  {journal} {\bibinfo
  {journal} {Physical Review}\ }\textbf {\bibinfo {volume} {37}},\ \bibinfo
  {pages} {405} (\bibinfo {year} {1931}{\natexlab{a}})}\BibitemShut {NoStop}%
\bibitem [{\citenamefont {Onsager}(1931{\natexlab{b}})}]{onsager1931}%
  \BibitemOpen
  \bibfield  {author} {\bibinfo {author} {\bibfnamefont {L.}~\bibnamefont
  {Onsager}},\ }\bibfield  {title} {\bibinfo {title} {Reciprocal {{Relations}}
  in {{Irreversible Processes}}. {{II}}.},\ }\href
  {https://doi.org/10.1103/PhysRev.38.2265} {\bibfield  {journal} {\bibinfo
  {journal} {Physical Review}\ }\textbf {\bibinfo {volume} {38}},\ \bibinfo
  {pages} {2265} (\bibinfo {year} {1931}{\natexlab{b}})}\BibitemShut {NoStop}%
\bibitem [{\citenamefont {Casimir}(1945)}]{casimir1945}%
  \BibitemOpen
  \bibfield  {author} {\bibinfo {author} {\bibfnamefont {H.~B.~G.}\
  \bibnamefont {Casimir}},\ }\bibfield  {title} {\bibinfo {title} {On
  {{Onsager}}'s {{Principle}} of {{Microscopic Reversibility}}},\ }\href
  {https://doi.org/10.1103/RevModPhys.17.343} {\bibfield  {journal} {\bibinfo
  {journal} {Reviews of Modern Physics}\ }\textbf {\bibinfo {volume} {17}},\
  \bibinfo {pages} {343} (\bibinfo {year} {1945})}\BibitemShut {NoStop}%
\bibitem [{\citenamefont {{Moncada-Villa}}\ \emph {et~al.}(2015)\citenamefont
  {{Moncada-Villa}}, \citenamefont {{Fern{\'a}ndez-Hurtado}}, \citenamefont
  {{Garc{\'i}a-Vidal}}, \citenamefont {{Garc{\'i}a-Mart{\'i}n}},\ and\
  \citenamefont {Cuevas}}]{moncada-villa2015}%
  \BibitemOpen
  \bibfield  {author} {\bibinfo {author} {\bibfnamefont {E.}~\bibnamefont
  {{Moncada-Villa}}}, \bibinfo {author} {\bibfnamefont {V.}~\bibnamefont
  {{Fern{\'a}ndez-Hurtado}}}, \bibinfo {author} {\bibfnamefont {F.~J.}\
  \bibnamefont {{Garc{\'i}a-Vidal}}}, \bibinfo {author} {\bibfnamefont
  {A.}~\bibnamefont {{Garc{\'i}a-Mart{\'i}n}}},\ and\ \bibinfo {author}
  {\bibfnamefont {J.~C.}\ \bibnamefont {Cuevas}},\ }\bibfield  {title}
  {\bibinfo {title} {Magnetic field control of near-field radiative heat
  transfer and the realization of highly tunable hyperbolic thermal emitters},\
  }\href {https://doi.org/10.1103/PhysRevB.92.125418} {\bibfield  {journal}
  {\bibinfo  {journal} {Physical Review B}\ }\textbf {\bibinfo {volume} {92}},\
  \bibinfo {pages} {125418} (\bibinfo {year} {2015})}\BibitemShut {NoStop}%
\bibitem [{\citenamefont {Zhu}\ and\ \citenamefont {Fan}(2014)}]{zhu2014}%
  \BibitemOpen
  \bibfield  {author} {\bibinfo {author} {\bibfnamefont {L.}~\bibnamefont
  {Zhu}}\ and\ \bibinfo {author} {\bibfnamefont {S.}~\bibnamefont {Fan}},\
  }\bibfield  {title} {\bibinfo {title} {Near-complete violation of detailed
  balance in thermal radiation},\ }\href
  {https://doi.org/10.1103/PhysRevB.90.220301} {\bibfield  {journal} {\bibinfo
  {journal} {Physical Review B}\ }\textbf {\bibinfo {volume} {90}},\ \bibinfo
  {pages} {220301} (\bibinfo {year} {2014})}\BibitemShut {NoStop}%
\bibitem [{\citenamefont {Silveirinha}(2017)}]{silveirinha2017}%
  \BibitemOpen
  \bibfield  {author} {\bibinfo {author} {\bibfnamefont {M.~G.}\ \bibnamefont
  {Silveirinha}},\ }\bibfield  {title} {\bibinfo {title} {Topological angular
  momentum and radiative heat transport in closed orbits},\ }\href
  {https://doi.org/10.1103/PhysRevB.95.115103} {\bibfield  {journal} {\bibinfo
  {journal} {Physical Review B}\ }\textbf {\bibinfo {volume} {95}},\ \bibinfo
  {pages} {115103} (\bibinfo {year} {2017})}\BibitemShut {NoStop}%
\bibitem [{\citenamefont {Abraham~Ekeroth}\ \emph {et~al.}(2018)\citenamefont
  {Abraham~Ekeroth}, \citenamefont {{Ben-Abdallah}}, \citenamefont {Cuevas},\
  and\ \citenamefont {{Garc{\'i}a-Mart{\'i}n}}}]{abrahamekeroth2018}%
  \BibitemOpen
  \bibfield  {author} {\bibinfo {author} {\bibfnamefont {R.~M.}\ \bibnamefont
  {Abraham~Ekeroth}}, \bibinfo {author} {\bibfnamefont {P.}~\bibnamefont
  {{Ben-Abdallah}}}, \bibinfo {author} {\bibfnamefont {J.~C.}\ \bibnamefont
  {Cuevas}},\ and\ \bibinfo {author} {\bibfnamefont {A.}~\bibnamefont
  {{Garc{\'i}a-Mart{\'i}n}}},\ }\bibfield  {title} {\bibinfo {title}
  {Anisotropic {{Thermal Magnetoresistance}} for an {{Active Control}} of
  {{Radiative Heat Transfer}}},\ }\href
  {https://doi.org/10.1021/acsphotonics.7b01223} {\bibfield  {journal}
  {\bibinfo  {journal} {ACS Photonics}\ }\textbf {\bibinfo {volume} {5}},\
  \bibinfo {pages} {705} (\bibinfo {year} {2018})}\BibitemShut {NoStop}%
\bibitem [{\citenamefont {Ott}\ \emph {et~al.}(2018)\citenamefont {Ott},
  \citenamefont {{Ben-Abdallah}},\ and\ \citenamefont {Biehs}}]{ott2018}%
  \BibitemOpen
  \bibfield  {author} {\bibinfo {author} {\bibfnamefont {A.}~\bibnamefont
  {Ott}}, \bibinfo {author} {\bibfnamefont {P.}~\bibnamefont
  {{Ben-Abdallah}}},\ and\ \bibinfo {author} {\bibfnamefont {S.-A.}\
  \bibnamefont {Biehs}},\ }\bibfield  {title} {\bibinfo {title} {Circular heat
  and momentum flux radiated by magneto-optical nanoparticles},\ }\href
  {https://doi.org/10.1103/PhysRevB.97.205414} {\bibfield  {journal} {\bibinfo
  {journal} {Physical Review B}\ }\textbf {\bibinfo {volume} {97}},\ \bibinfo
  {pages} {205414} (\bibinfo {year} {2018})}\BibitemShut {NoStop}%
\bibitem [{\citenamefont {Ott}\ \emph {et~al.}(2019)\citenamefont {Ott},
  \citenamefont {Messina}, \citenamefont {{Ben-Abdallah}},\ and\ \citenamefont
  {Biehs}}]{ott2019a}%
  \BibitemOpen
  \bibfield  {author} {\bibinfo {author} {\bibfnamefont {A.}~\bibnamefont
  {Ott}}, \bibinfo {author} {\bibfnamefont {R.}~\bibnamefont {Messina}},
  \bibinfo {author} {\bibfnamefont {P.}~\bibnamefont {{Ben-Abdallah}}},\ and\
  \bibinfo {author} {\bibfnamefont {S.-A.}\ \bibnamefont {Biehs}},\ }\bibfield
  {title} {\bibinfo {title} {Radiative thermal diode driven by nonreciprocal
  surface waves},\ }\href {https://doi.org/10.1063/1.5093626} {\bibfield
  {journal} {\bibinfo  {journal} {Applied Physics Letters}\ }\textbf {\bibinfo
  {volume} {114}},\ \bibinfo {pages} {163105} (\bibinfo {year}
  {2019})}\BibitemShut {NoStop}%
\bibitem [{\citenamefont {Zhao}\ \emph {et~al.}(2019)\citenamefont {Zhao},
  \citenamefont {Shi}, \citenamefont {Wang}, \citenamefont {Zhao},
  \citenamefont {Zhao},\ and\ \citenamefont {Fan}}]{zhao2019a}%
  \BibitemOpen
  \bibfield  {author} {\bibinfo {author} {\bibfnamefont {B.}~\bibnamefont
  {Zhao}}, \bibinfo {author} {\bibfnamefont {Y.}~\bibnamefont {Shi}}, \bibinfo
  {author} {\bibfnamefont {J.}~\bibnamefont {Wang}}, \bibinfo {author}
  {\bibfnamefont {Z.}~\bibnamefont {Zhao}}, \bibinfo {author} {\bibfnamefont
  {N.}~\bibnamefont {Zhao}},\ and\ \bibinfo {author} {\bibfnamefont
  {S.}~\bibnamefont {Fan}},\ }\bibfield  {title} {\bibinfo {title}
  {Near-complete violation of {{Kirchhoff}}'s law of thermal radiation with a
  0.3 {{T}} magnetic field},\ }\href {https://doi.org/10.1364/OL.44.004203}
  {\bibfield  {journal} {\bibinfo  {journal} {Optics Letters}\ }\textbf
  {\bibinfo {volume} {44}},\ \bibinfo {pages} {4203} (\bibinfo {year}
  {2019})}\BibitemShut {NoStop}%
\bibitem [{\citenamefont {Fan}\ \emph {et~al.}(2020)\citenamefont {Fan},
  \citenamefont {Guo}, \citenamefont {Papadakis}, \citenamefont {Zhao},
  \citenamefont {Zhao}, \citenamefont {Buddhiraju}, \citenamefont {Orenstein},\
  and\ \citenamefont {Fan}}]{fan2020a}%
  \BibitemOpen
  \bibfield  {author} {\bibinfo {author} {\bibfnamefont {L.}~\bibnamefont
  {Fan}}, \bibinfo {author} {\bibfnamefont {Y.}~\bibnamefont {Guo}}, \bibinfo
  {author} {\bibfnamefont {G.~T.}\ \bibnamefont {Papadakis}}, \bibinfo {author}
  {\bibfnamefont {B.}~\bibnamefont {Zhao}}, \bibinfo {author} {\bibfnamefont
  {Z.}~\bibnamefont {Zhao}}, \bibinfo {author} {\bibfnamefont {S.}~\bibnamefont
  {Buddhiraju}}, \bibinfo {author} {\bibfnamefont {M.}~\bibnamefont
  {Orenstein}},\ and\ \bibinfo {author} {\bibfnamefont {S.}~\bibnamefont
  {Fan}},\ }\bibfield  {title} {\bibinfo {title} {Nonreciprocal radiative heat
  transfer between two planar bodies},\ }\href
  {https://doi.org/10.1103/PhysRevB.101.085407} {\bibfield  {journal} {\bibinfo
   {journal} {Physical Review B}\ }\textbf {\bibinfo {volume} {101}},\ \bibinfo
  {pages} {085407} (\bibinfo {year} {2020})}\BibitemShut {NoStop}%
\bibitem [{\citenamefont {Zhao}\ \emph {et~al.}(2020)\citenamefont {Zhao},
  \citenamefont {Guo}, \citenamefont {Garcia}, \citenamefont {Narang},\ and\
  \citenamefont {Fan}}]{zhao2020e}%
  \BibitemOpen
  \bibfield  {author} {\bibinfo {author} {\bibfnamefont {B.}~\bibnamefont
  {Zhao}}, \bibinfo {author} {\bibfnamefont {C.}~\bibnamefont {Guo}}, \bibinfo
  {author} {\bibfnamefont {C.~A.~C.}\ \bibnamefont {Garcia}}, \bibinfo {author}
  {\bibfnamefont {P.}~\bibnamefont {Narang}},\ and\ \bibinfo {author}
  {\bibfnamefont {S.}~\bibnamefont {Fan}},\ }\bibfield  {title} {\bibinfo
  {title} {Axion-{{Field}}-{{Enabled Nonreciprocal Thermal Radiation}} in
  {{Weyl Semimetals}}},\ }\href {https://doi.org/10.1021/acs.nanolett.9b05179}
  {\bibfield  {journal} {\bibinfo  {journal} {Nano Letters}\ }\textbf {\bibinfo
  {volume} {20}},\ \bibinfo {pages} {1923} (\bibinfo {year}
  {2020})}\BibitemShut {NoStop}%
\bibitem [{\citenamefont {Tsurimaki}\ \emph {et~al.}(2020)\citenamefont
  {Tsurimaki}, \citenamefont {Qian}, \citenamefont {Pajovic}, \citenamefont
  {Han}, \citenamefont {Li},\ and\ \citenamefont {Chen}}]{tsurimaki2020}%
  \BibitemOpen
  \bibfield  {author} {\bibinfo {author} {\bibfnamefont {Y.}~\bibnamefont
  {Tsurimaki}}, \bibinfo {author} {\bibfnamefont {X.}~\bibnamefont {Qian}},
  \bibinfo {author} {\bibfnamefont {S.}~\bibnamefont {Pajovic}}, \bibinfo
  {author} {\bibfnamefont {F.}~\bibnamefont {Han}}, \bibinfo {author}
  {\bibfnamefont {M.}~\bibnamefont {Li}},\ and\ \bibinfo {author}
  {\bibfnamefont {G.}~\bibnamefont {Chen}},\ }\bibfield  {title} {\bibinfo
  {title} {Large nonreciprocal absorption and emission of radiation in type-{{I
  Weyl}} semimetals with time reversal symmetry breaking},\ }\href
  {https://doi.org/10.1103/PhysRevB.101.165426} {\bibfield  {journal} {\bibinfo
   {journal} {Physical Review B}\ }\textbf {\bibinfo {volume} {101}},\ \bibinfo
  {pages} {165426} (\bibinfo {year} {2020})}\BibitemShut {NoStop}%
\bibitem [{\citenamefont {Ott}\ \emph {et~al.}(2020)\citenamefont {Ott},
  \citenamefont {Biehs},\ and\ \citenamefont {{Ben-Abdallah}}}]{ott2020a}%
  \BibitemOpen
  \bibfield  {author} {\bibinfo {author} {\bibfnamefont {A.}~\bibnamefont
  {Ott}}, \bibinfo {author} {\bibfnamefont {S.-A.}\ \bibnamefont {Biehs}},\
  and\ \bibinfo {author} {\bibfnamefont {P.}~\bibnamefont {{Ben-Abdallah}}},\
  }\bibfield  {title} {\bibinfo {title} {Anomalous photon thermal {{Hall}}
  effect},\ }\href {https://doi.org/10.1103/PhysRevB.101.241411} {\bibfield
  {journal} {\bibinfo  {journal} {Physical Review B}\ }\textbf {\bibinfo
  {volume} {101}},\ \bibinfo {pages} {241411} (\bibinfo {year}
  {2020})}\BibitemShut {NoStop}%
\bibitem [{\citenamefont {Zhu}\ and\ \citenamefont {Fan}(2016)}]{zhu2016}%
  \BibitemOpen
  \bibfield  {author} {\bibinfo {author} {\bibfnamefont {L.}~\bibnamefont
  {Zhu}}\ and\ \bibinfo {author} {\bibfnamefont {S.}~\bibnamefont {Fan}},\
  }\bibfield  {title} {\bibinfo {title} {Persistent {{Directional Current}} at
  {{Equilibrium}} in {{Nonreciprocal Many}}-{{Body Near Field Electromagnetic
  Heat Transfer}}},\ }\href {https://doi.org/10.1103/PhysRevLett.117.134303}
  {\bibfield  {journal} {\bibinfo  {journal} {Physical Review Letters}\
  }\textbf {\bibinfo {volume} {117}},\ \bibinfo {pages} {134303} (\bibinfo
  {year} {2016})}\BibitemShut {NoStop}%
\bibitem [{\citenamefont {{Ben-Abdallah}}(2016)}]{ben-abdallah2016}%
  \BibitemOpen
  \bibfield  {author} {\bibinfo {author} {\bibfnamefont {P.}~\bibnamefont
  {{Ben-Abdallah}}},\ }\bibfield  {title} {\bibinfo {title} {Photon {{Thermal
  Hall Effect}}},\ }\href {https://doi.org/10.1103/PhysRevLett.116.084301}
  {\bibfield  {journal} {\bibinfo  {journal} {Physical Review Letters}\
  }\textbf {\bibinfo {volume} {116}},\ \bibinfo {pages} {084301} (\bibinfo
  {year} {2016})}\BibitemShut {NoStop}%
\bibitem [{\citenamefont {Guo}\ \emph {et~al.}(2019)\citenamefont {Guo},
  \citenamefont {Guo},\ and\ \citenamefont {Fan}}]{guo2019a}%
  \BibitemOpen
  \bibfield  {author} {\bibinfo {author} {\bibfnamefont {C.}~\bibnamefont
  {Guo}}, \bibinfo {author} {\bibfnamefont {Y.}~\bibnamefont {Guo}},\ and\
  \bibinfo {author} {\bibfnamefont {S.}~\bibnamefont {Fan}},\ }\bibfield
  {title} {\bibinfo {title} {Relation between photon thermal {{Hall}} effect
  and persistent heat current in nonreciprocal radiative heat transfer},\
  }\href {https://doi.org/10.1103/PhysRevB.100.205416} {\bibfield  {journal}
  {\bibinfo  {journal} {Physical Review B}\ }\textbf {\bibinfo {volume}
  {100}},\ \bibinfo {pages} {205416} (\bibinfo {year} {2019})}\BibitemShut
  {NoStop}%
\bibitem [{\citenamefont {{Ben-Abdallah}}\ \emph {et~al.}(2011)\citenamefont
  {{Ben-Abdallah}}, \citenamefont {Biehs},\ and\ \citenamefont
  {Joulain}}]{ben-abdallah2011}%
  \BibitemOpen
  \bibfield  {author} {\bibinfo {author} {\bibfnamefont {P.}~\bibnamefont
  {{Ben-Abdallah}}}, \bibinfo {author} {\bibfnamefont {S.-A.}\ \bibnamefont
  {Biehs}},\ and\ \bibinfo {author} {\bibfnamefont {K.}~\bibnamefont
  {Joulain}},\ }\bibfield  {title} {\bibinfo {title} {Many-{{Body Radiative
  Heat Transfer Theory}}},\ }\href
  {https://doi.org/10.1103/PhysRevLett.107.114301} {\bibfield  {journal}
  {\bibinfo  {journal} {Physical Review Letters}\ }\textbf {\bibinfo {volume}
  {107}},\ \bibinfo {pages} {114301} (\bibinfo {year} {2011})}\BibitemShut
  {NoStop}%
\bibitem [{\citenamefont {Khandekar}\ and\ \citenamefont
  {Jacob}(2019)}]{khandekar2019b}%
  \BibitemOpen
  \bibfield  {author} {\bibinfo {author} {\bibfnamefont {C.}~\bibnamefont
  {Khandekar}}\ and\ \bibinfo {author} {\bibfnamefont {Z.}~\bibnamefont
  {Jacob}},\ }\bibfield  {title} {\bibinfo {title} {Circularly {{Polarized
  Thermal Radiation From Nonequilibrium Coupled Antennas}}},\ }\href
  {https://doi.org/10.1103/PhysRevApplied.12.014053} {\bibfield  {journal}
  {\bibinfo  {journal} {Physical Review Applied}\ }\textbf {\bibinfo {volume}
  {12}},\ \bibinfo {pages} {014053} (\bibinfo {year} {2019})}\BibitemShut
  {NoStop}%
\bibitem [{\citenamefont {Tretyakov}\ \emph {et~al.}(2002)\citenamefont
  {Tretyakov}, \citenamefont {Sihvola},\ and\ \citenamefont
  {Jancewicz}}]{tretyakov2002}%
  \BibitemOpen
  \bibfield  {author} {\bibinfo {author} {\bibfnamefont {S.}~\bibnamefont
  {Tretyakov}}, \bibinfo {author} {\bibfnamefont {A.}~\bibnamefont {Sihvola}},\
  and\ \bibinfo {author} {\bibfnamefont {B.}~\bibnamefont {Jancewicz}},\
  }\bibfield  {title} {\bibinfo {title} {Onsager-{{Casimir Principle}} and the
  {{Constitutive Relations}} of {{Bi}}-{{Anisotropic Media}}},\ }\href
  {https://doi.org/10.1163/156939302X00453} {\bibfield  {journal} {\bibinfo
  {journal} {Journal of Electromagnetic Waves and Applications}\ }\textbf
  {\bibinfo {volume} {16}},\ \bibinfo {pages} {573} (\bibinfo {year}
  {2002})}\BibitemShut {NoStop}%
\bibitem [{\citenamefont {Zhu}\ \emph {et~al.}(2018)\citenamefont {Zhu},
  \citenamefont {Guo},\ and\ \citenamefont {Fan}}]{zhu2018}%
  \BibitemOpen
  \bibfield  {author} {\bibinfo {author} {\bibfnamefont {L.}~\bibnamefont
  {Zhu}}, \bibinfo {author} {\bibfnamefont {Y.}~\bibnamefont {Guo}},\ and\
  \bibinfo {author} {\bibfnamefont {S.}~\bibnamefont {Fan}},\ }\bibfield
  {title} {\bibinfo {title} {Theory of many-body radiative heat transfer
  without the constraint of reciprocity},\ }\href
  {https://doi.org/10.1103/PhysRevB.97.094302} {\bibfield  {journal} {\bibinfo
  {journal} {Physical Review B}\ }\textbf {\bibinfo {volume} {97}},\ \bibinfo
  {pages} {094302} (\bibinfo {year} {2018})}\BibitemShut {NoStop}%
\bibitem [{\citenamefont {Bradley}\ and\ \citenamefont
  {Cracknell}(2010)}]{bradley2010}%
  \BibitemOpen
  \bibfield  {author} {\bibinfo {author} {\bibfnamefont {C.~J.}\ \bibnamefont
  {Bradley}}\ and\ \bibinfo {author} {\bibfnamefont {A.~P.}\ \bibnamefont
  {Cracknell}},\ }\href@noop {} {\emph {\bibinfo {title} {The Mathematical
  Theory of Symmetry in Solids: Representation Theory for Point Groups and
  Space Groups}}}\ (\bibinfo  {publisher} {{Clarendon Press}},\ \bibinfo
  {address} {{Oxford}},\ \bibinfo {year} {2010})\BibitemShut {NoStop}%
\bibitem [{\citenamefont {Hamermesh}(1989)}]{hamermesh1989}%
  \BibitemOpen
  \bibfield  {author} {\bibinfo {author} {\bibfnamefont {M.}~\bibnamefont
  {Hamermesh}},\ }\href@noop {} {\emph {\bibinfo {title} {Group Theory and Its
  Application to Physical Problems}}}\ (\bibinfo  {publisher} {{Dover
  Publications}},\ \bibinfo {address} {{New York}},\ \bibinfo {year}
  {1989})\BibitemShut {NoStop}%
\bibitem [{\citenamefont {Dresselhaus}\ \emph {et~al.}(2010)\citenamefont
  {Dresselhaus}, \citenamefont {Dresselhaus},\ and\ \citenamefont
  {Jorio}}]{dresselhaus2010}%
  \BibitemOpen
  \bibfield  {author} {\bibinfo {author} {\bibfnamefont {M.~S.}\ \bibnamefont
  {Dresselhaus}}, \bibinfo {author} {\bibfnamefont {G.}~\bibnamefont
  {Dresselhaus}},\ and\ \bibinfo {author} {\bibfnamefont {A.}~\bibnamefont
  {Jorio}},\ }\href@noop {} {\emph {\bibinfo {title} {Group Theory: Application
  to the Physics of Condensed Matter}}}\ (\bibinfo  {publisher} {{Springer}},\
  \bibinfo {address} {{Berlin; Heidelberg}},\ \bibinfo {year}
  {2010})\BibitemShut {NoStop}%
\bibitem [{\citenamefont {{Jin Au Kong}}(1972)}]{jinaukong1972}%
  \BibitemOpen
  \bibfield  {author} {\bibinfo {author} {\bibnamefont {{Jin Au Kong}}},\
  }\bibfield  {title} {\bibinfo {title} {Theorems of bianisotropic media},\
  }\href {https://doi.org/10.1109/PROC.1972.8851} {\bibfield  {journal}
  {\bibinfo  {journal} {Proceedings of the IEEE}\ }\textbf {\bibinfo {volume}
  {60}},\ \bibinfo {pages} {1036} (\bibinfo {year} {1972})}\BibitemShut
  {NoStop}%
\bibitem [{\citenamefont {Palik}(1998)}]{palik1998handbook}%
  \BibitemOpen
  \bibfield  {author} {\bibinfo {author} {\bibfnamefont {E.~D.}\ \bibnamefont
  {Palik}},\ }\href@noop {} {\emph {\bibinfo {title} {{Handbook of optical
  constants of solids}}}},\ Vol.~\bibinfo {volume} {3}\ (\bibinfo  {publisher}
  {Academic press},\ \bibinfo {year} {1998})\BibitemShut {NoStop}%
\bibitem [{\citenamefont {Rumsey}(1954)}]{rumsey1954}%
  \BibitemOpen
  \bibfield  {author} {\bibinfo {author} {\bibfnamefont {V.~H.}\ \bibnamefont
  {Rumsey}},\ }\bibfield  {title} {\bibinfo {title} {Reaction {{Concept}} in
  {{Electromagnetic Theory}}},\ }\href
  {https://doi.org/10.1103/PhysRev.94.1483} {\bibfield  {journal} {\bibinfo
  {journal} {Physical Review}\ }\textbf {\bibinfo {volume} {94}},\ \bibinfo
  {pages} {1483} (\bibinfo {year} {1954})}\BibitemShut {NoStop}%
\bibitem [{\citenamefont {Villeneuve}\ and\ \citenamefont
  {Harrington}(1958)}]{villeneuve1958}%
  \BibitemOpen
  \bibfield  {author} {\bibinfo {author} {\bibfnamefont {A.}~\bibnamefont
  {Villeneuve}}\ and\ \bibinfo {author} {\bibfnamefont {R.}~\bibnamefont
  {Harrington}},\ }\bibfield  {title} {\bibinfo {title} {Reciprocity
  {{Relationships}} for {{Gyrotropic Media}}},\ }\href
  {https://doi.org/10.1109/TMTT.1958.1124563} {\bibfield  {journal} {\bibinfo
  {journal} {IRE Transactions on Microwave Theory and Techniques}\ }\textbf
  {\bibinfo {volume} {6}},\ \bibinfo {pages} {308} (\bibinfo {year}
  {1958})}\BibitemShut {NoStop}%
\bibitem [{\citenamefont {Kr{\"u}ger}\ \emph {et~al.}(2012)\citenamefont
  {Kr{\"u}ger}, \citenamefont {Bimonte}, \citenamefont {Emig},\ and\
  \citenamefont {Kardar}}]{kruger2012}%
  \BibitemOpen
  \bibfield  {author} {\bibinfo {author} {\bibfnamefont {M.}~\bibnamefont
  {Kr{\"u}ger}}, \bibinfo {author} {\bibfnamefont {G.}~\bibnamefont {Bimonte}},
  \bibinfo {author} {\bibfnamefont {T.}~\bibnamefont {Emig}},\ and\ \bibinfo
  {author} {\bibfnamefont {M.}~\bibnamefont {Kardar}},\ }\bibfield  {title}
  {\bibinfo {title} {Trace formulas for nonequilibrium {{Casimir}}
  interactions, heat radiation, and heat transfer for arbitrary objects},\
  }\href {https://doi.org/10.1103/PhysRevB.86.115423} {\bibfield  {journal}
  {\bibinfo  {journal} {Physical Review B}\ }\textbf {\bibinfo {volume} {86}},\
  \bibinfo {pages} {115423} (\bibinfo {year} {2012})}\BibitemShut {NoStop}%
\end{thebibliography}

%

\end{document}